\documentclass[11pt,twoside]{article}
\usepackage{asp2014}

\aspSuppressVolSlug
\resetcounters

\begin{document}

\title{Understanding Systematics in ZZ Ceti Model Fitting to Enable Differential Seismology}
\author{J.T.~Fuchs,$^1$ Bart~H.~Dunlap,$^1$ J.~C.~Clemens,$^1$ J.~A.~Meza,$^1$ E.~Dennihy,$^1$ and D.~Koester$^2$
\affil{$^1$University of North Carolina at Chapel Hill, Chapel Hill, NC, USA; \email{joshfuchs@unc.edu}}
\affil{$^2$Institut f\"{u}r Theoretische Physik und Astrophysik, University of Kiel, D-24098 Kiel, Germany}}

\paperauthor{J.T.~Fuchs}{joshfuchs@unc.edu}{0000-0003-2775-3304}{University of North Carolina at Chapel Hill}{Department of Physics and Astronomy}{Chapel Hill}{NC}{27599}{USA}
\paperauthor{Bart~H.~Dunlap}{bhdunlap@physics.unc.edu}{}{University of North Carolina at Chapel Hill}{Department of Physics and Astronomy}{Chapel Hill}{NC}{27599}{USA}
\paperauthor{ J.~C.~Clemens}{clemens@physics.unc.edu}{}{University of North Carolina at Chapel Hill}{Department of Physics and Astronomy}{Chapel Hill}{NC}{27599}{USA}
\paperauthor{ J.~A.~Meza}{}{}{University of North Carolina at Chapel Hill}{Department of Physics and Astronomy}{Chapel Hill}{NC}{27599}{USA}
\paperauthor{E.~Dennihy}{edennihy@live.unc.edu}{}{University of North Carolina at Chapel Hill}{Department of Physics and Astronomy}{Chapel Hill}{NC}{27599}{USA}
\paperauthor{D.~Koester}{koester@astrophysik.uni-kiel.de}{}{University of Kiel}{Institut f\"{u}r Theoretische Physik und Astrophysik}{Kiel}{}{D-24098}{Germany}

\begin{abstract}
We are conducting a large spectroscopic survey of over 130 Southern ZZ Cetis with the Goodman Spectrograph on the SOAR Telescope. Because it employs a single instrument with high UV throughput, this survey will both improve the signal-to-noise of the sample of SDSS ZZ Cetis and provide a uniform dataset for model comparison. We are paying special attention to systematics in the spectral fitting and quantify three of those systematics here. We show that relative positions in the $\log{g}$-$T_{\rm eff}$ plane are consistent for these three systematics.
\end{abstract}

\section{Introduction}
We have undertaken an observing project to collect high signal-to-noise spectroscopy of more than 130 hydrogen-atmosphere pulsating white dwarfs, also known as ZZ Cetis. This includes all ZZ Cetis south of $+10^{\circ}$ declination as well as those observed by the {\em K2} mission \citep{2014PASP..126..398H}. We are observing all targets with the Goodman Spectrograph on the SOAR Telescope \citep{2004SPIE.5492..331C}. While many of these ZZ Cetis have been spectroscopically observed by the Sloan Digital Sky Survey (SDSS, \citealt{2000AJ....120.1579Y}), most of those are of low signal-to-noise resulting in large $T_{\rm eff}$ and $\log{g}$ uncertainties. Our spectra will have a signal-to-noise of at least 80. The homogeneity of this sample allows us to carefully explore systematics in the determination of atmospheric parameters. Minimizing these systematics will help determine relative positions of each ZZ Ceti in the $\log{g}$-$T_{\rm eff}$ plane.

To determine the atmospheric parameters, we use the spectroscopic method as described and improved in \citet{1992ApJ...394..228B}, \citet{1995ApJ...449..258B}, and \citet{2005ApJS..156...47L}. In short, this procedure normalizes spectral lines by first fitting pseudogaussians to each of the Balmer lines. This helps determine both the center of the line and normalization points at fixed wavelengths. Each line is then individually normalized using the best-fitting profile. These normalized line profiles are compared to model spectra from \citet{2010MmSAI..81..921K}. The model spectra are first convolved to match the observed resolution. The individual Balmer lines of the convolved model spectra are then each normalized using the same normalization wavelengths used for the observed spectrum. Finally, we compare our normalized spectrum to a grid of normalized model spectra to determine the atmospheric parameters, $T_{\rm eff}$ and $\log{g}$. 

The formal uncertainties on these fitted parameters are determined by the signal-to-noise of the spectrum (see \citealt{2005ApJ...631.1100G}). However, external factors may account for larger errors. To properly calibrate our sample of ZZ Ceti spectra, we seek to quantify these systematics in order to determine the relative positions of the stars. This is important for differential seismology, in which differences in mode periods of same $\ell$ and $k$ are compared to differences between models of stars with similar $T_{\rm eff}$ and $\log{g}$. Here we present initial analysis showing the effect of three systematics on the final atmospheric parameters.

\section{Systematics}
We have so far investigated three choices that must be made to estimate $T_{\rm eff}$ and $\log{g}$. First, we will look at differences caused by flux calibration with different standard stars. Second, we will investigate the importance of convolving the models with the correct resolution, which is set by the seeing and instrument profile and measured from the data. Third, we will choose different continuum normalization points for each line. We will illustrate each systematic using an observation of the hot ZZ Ceti star GD 165 (V = 14.3 mag). However, the four stars discussed in Section \ref{sec:ordering} all show the same behavior for each systematic.

\articlefigure[width=0.8\textwidth]{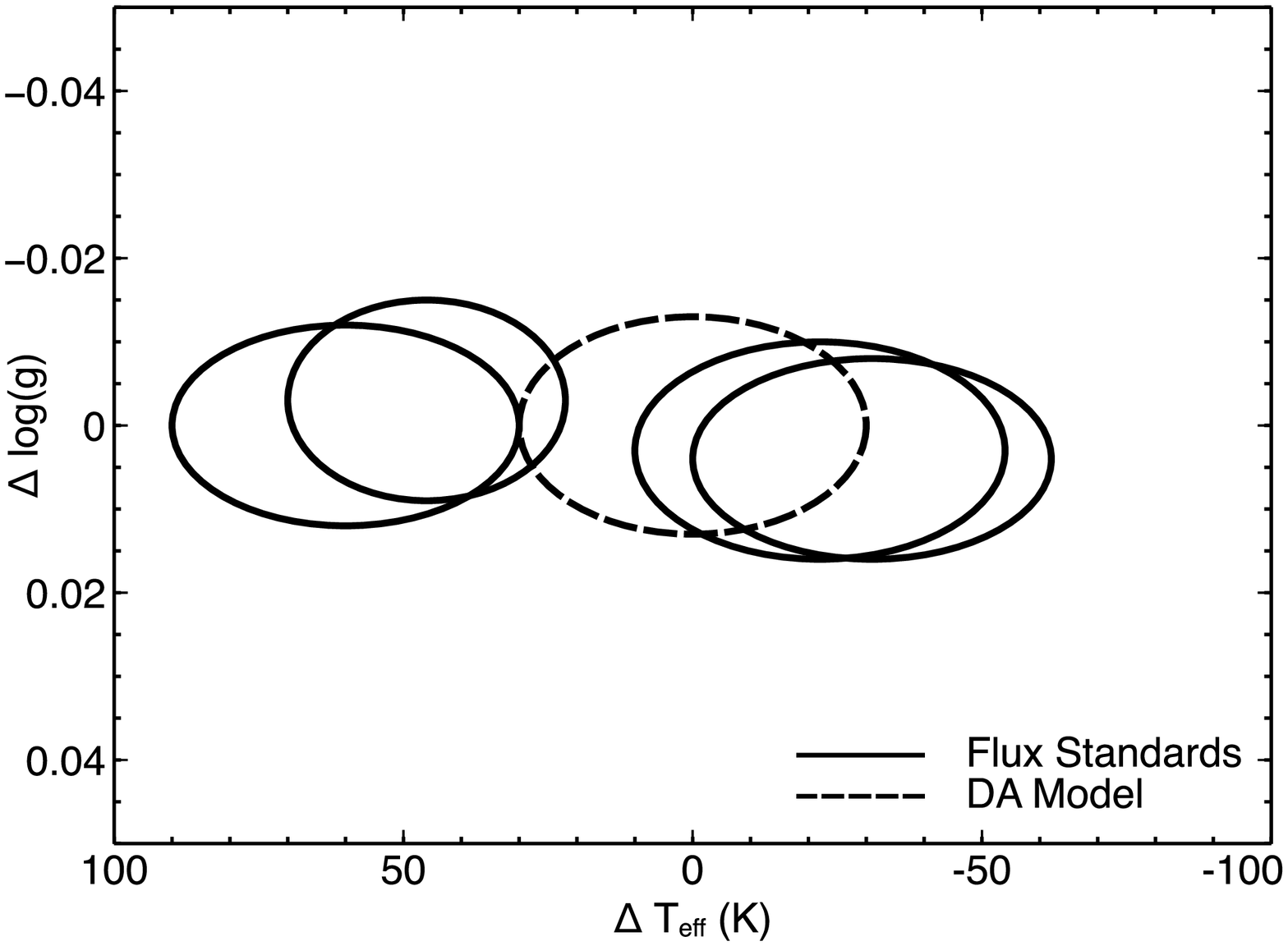}{fig:flux_standards}{Using different flux standards changes the final determined atmospheric parameters. Different flux standards from the same night are shown as solid ellipses. We also use a white dwarf model for relative flux calibration, shown as a dashed ellipse. We show how $T_{\rm eff}$ and $\log{g}$ change for each standard star relative to using the model. We see no effect correlated with airmass or observation time.}

\subsection{Flux Standard}
First, we investigate how choosing different standard stars for flux calibration changes our final atmospheric parameters. Our observations use a 3$''$ slit and have typical seeing of 1$''$-1.5$''$, so we lose very little light at the slit and therefore have good flux calibration. We observed four different ESO standards at different times and airmasses. For further comparison, we also used a model for relative flux calibration. We used a model at 12,000 K and $\log{g}$ = 8.0, though the results of the line fitting are not sensitive to the exact model used. We show the results in Figure \ref{fig:flux_standards}. Different choices of flux standards give what we presume are stochastic variations. There is no trend with airmass or time of observation. The results vary by about 125 K and 0.01 in $\log{g}$.

\subsection{Model Convolution}
We next investigate the importance of convolving the models to correctly match the resolution of the spectrum, which we measure by computing the FWHM of the spectrum in the spatial direction. We tested values 20 and 40 percent larger and smaller than our measured resolution and show our results in Figure \ref{fig:seeing}. Convolving the models with a lower resolution than the spectrum leads to cooler and less massive fits. Convolving the models with a higher resolution than the spectrum leads to hotter and more massive fits. The range in the atmospheric parameters is 200 K and 0.04 in $\log{g}$.

\articlefigure[width=0.8\textwidth]{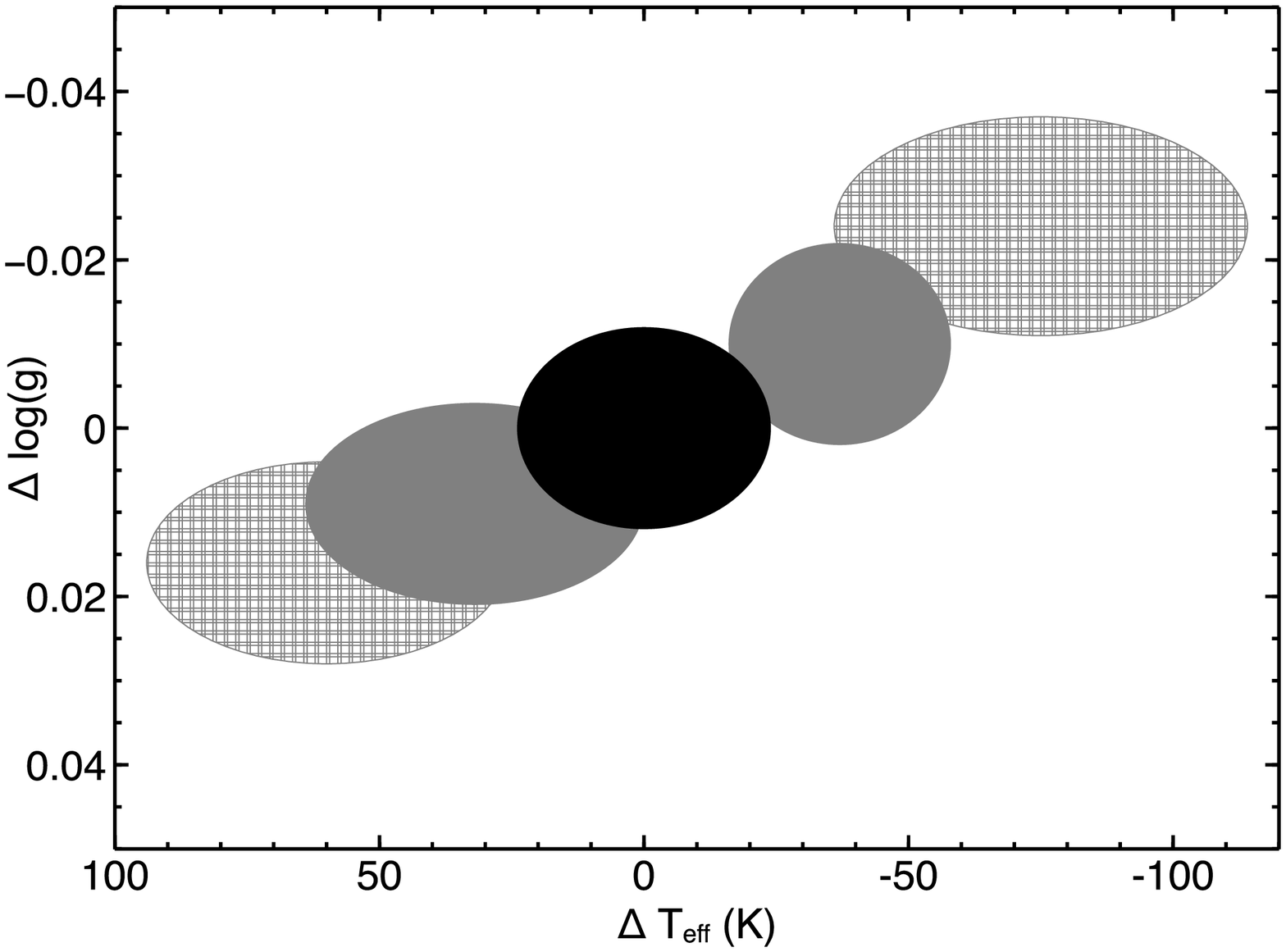}{fig:seeing}{Convolving the models with different resolutions than the observed spectrum also changes the final atmospheric parameters. Black is the measured resolution from the spectrum. Solid grey is convolving with resolutions 20 percent larger and smaller. Hatched grey is convolving with resolutions 40 percent larger and smaller. Convolving the models with a lower resolution than the spectrum leads to cooler and less massive fits. Convolving the models with a higher resolution than the spectrum leads to hotter and more massive fits.}

The explanation for this effect is that as we convolve the models with different resolutions, we are slightly changing the width of the lines. Erroneously low resolution causes the model Balmer lines to become broader and shallower, which leads to an inference of  cooler, less massive stars.

\subsection{Normalization Points}
Finally, we test the effect of changing the continuum normalization points that are defined in \citet{2005ApJS..156...47L}. We narrowed and widened the location of our normalization points so that each line width we compared was 15 percent smaller or larger than \citet{2005ApJS..156...47L}. We show the result on the atmospheric parameters in Figure \ref{fig:normalization}. The range of atmospheric parameters in this case is 350 K and 0.06 in $\log{g}$. 

The typical normalization points are chosen to be close to the continuum value between the Balmer lines. As we change the normalization points, we are changing the amount of information in the lines. Widening the lines add marginally more information to the line depth. Narrowing the lines causes the loss of more information, leading to a larger change in $T_{\rm eff}$ and $\log{g}$. We will continue this investigation to determine which choice yields the most accurate atmospheric parameters.

\articlefigure[width=0.8\textwidth]{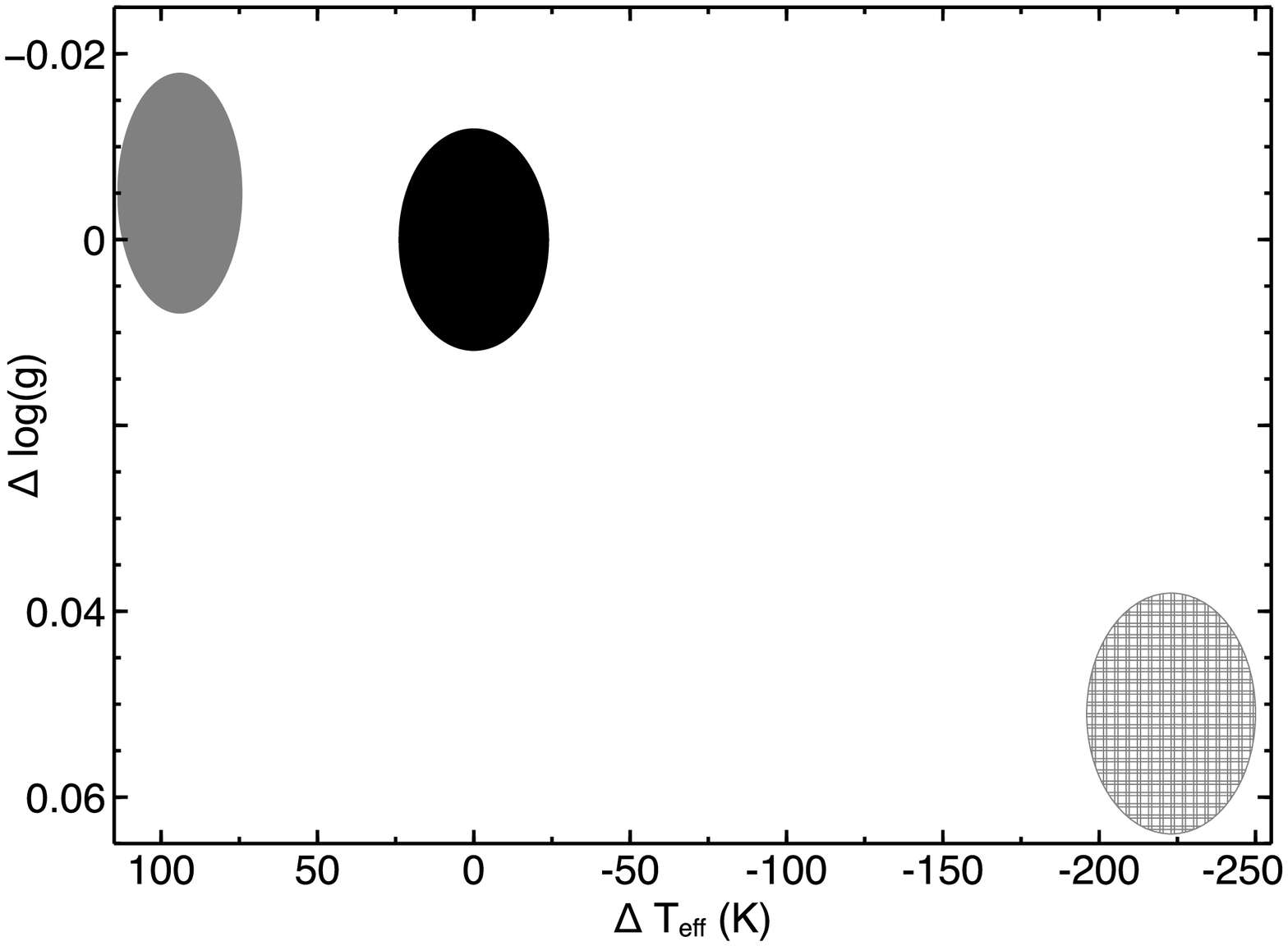}{fig:normalization}{Different normalization points likewise give different final atmospheric parameters. Solid black shows using the line widths as defined in \citet{2005ApJS..156...47L}. Solid grey results from compared line widths 15 percent wider while hatched grey results from compared line widths 15 percent narrower.}

\section{Relative Ordering}
\label{sec:ordering}
Even if our absolute atmospheric parameters are afflicted by systematics, initial results show that the relative orderings are more secure. To investigate the relative ordering between stars, we look at results from four of the brightest ZZ Cetis: L19-2, GD 165, GD 133, and WD 1150-153. For the three systematics we have discussed above, the relative ordering between these four ZZ Cetis stays the same, as shown in Figure \ref{fig:4horsemen}. This relative ordering is important for the differential seismology work our group is pursuing.

\articlefigure[width=\textwidth]{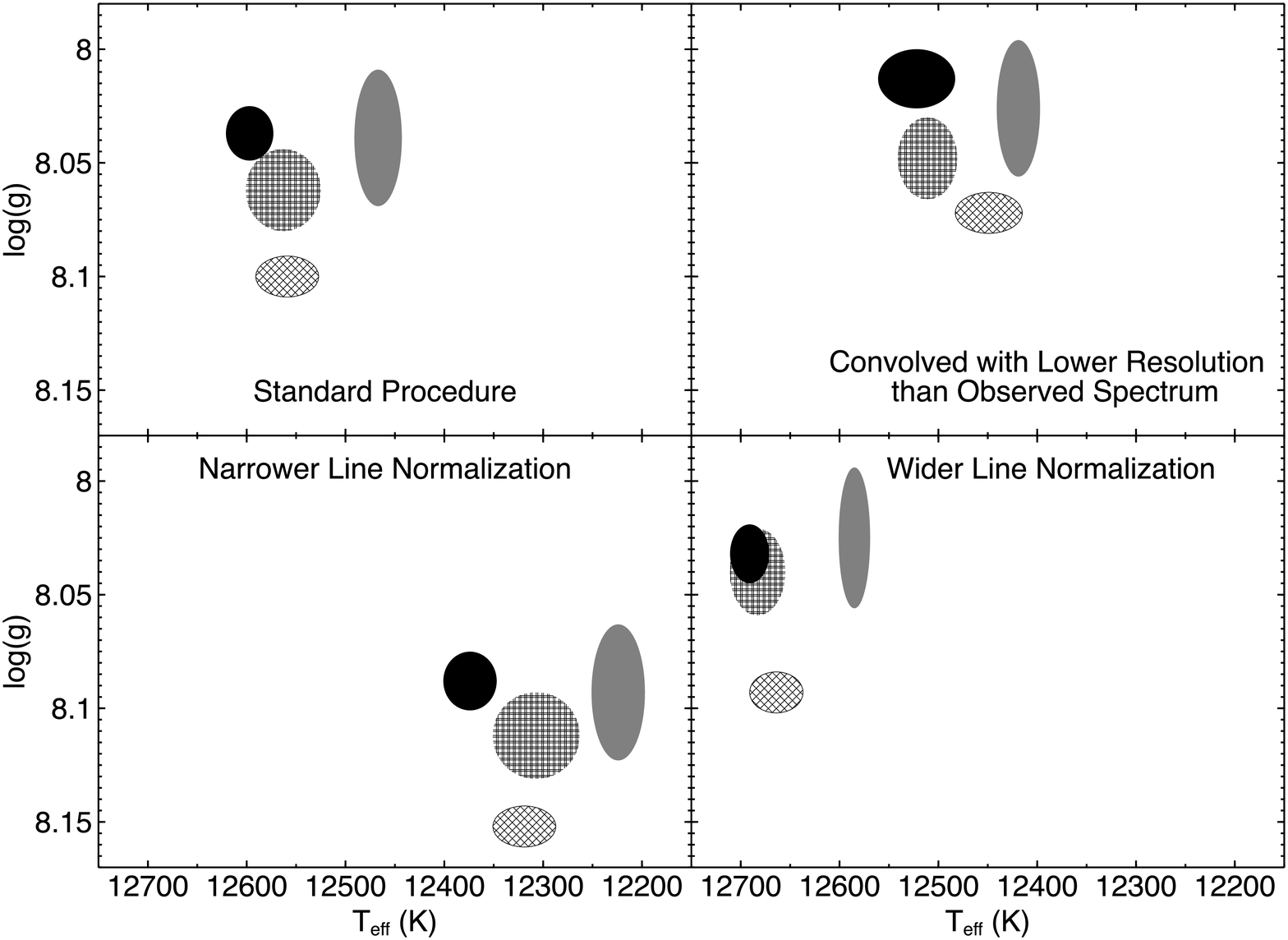}{fig:4horsemen}{The relative ordering between four bright ZZ Cetis is the same regardless of the systematics shown above. Each panel shows results from changing different aspects of the data reduction and fitting procedure discussed above.}

\section{Conclusions}
We have presented initial results from our investigation into systematics affecting atmospheric parameter determination of ZZ Cetis using the spectroscopic method. We have found systematics can change $T_{\rm eff}$ by up to 350 K and $\log{g}$ by up to 0.06 dex. The formal uncertainties from our fitting are around 50 K and 0.02 dex. While the systematics explored here are larger than the formal errors, for relative positioning each of these systematics can be accounted for by using consistent normalization points, the correct resolution when convolving the models, and relative flux calibration using a DA model. Future work will continue to investigate other systematics in the reduction and fitting process. However, the relative ordering between various stars stays the same for the systematics we have investigated. We will soon have the relative order for the more than 130 ZZ Cetis observed in our survey.

\acknowledgements The authors thank P. Bergeron for helpful discussions. JTF, JCC, JAM, and ED acknowledge support from the National Science Foundation under award AST-1413001. Based on observations obtained at the Southern Astrophysical Research (SOAR) telescope, which is a joint project of the Minist\'{e}rio da Ci\^{e}ncia, Tecnologia, e Inova\c{c}\~{a}o (MCTI) da Rep\'{u}blica Federativa do Brasil, the U.S. National Optical Astronomy Observatory (NOAO), the University of North Carolina at Chapel Hill (UNC), and Michigan State University (MSU). Figures were made with Veusz, a free scientific plotting package written by Jeremy Sanders. Verusz can be found at http://home.gna.org/veusz/.

\def\memsai{Mem. Soc. Astron. Ital.} 
\def\procspie{Proc. SPIE Conf. Ser.} 

\end{document}